# Characterization of Single-Walled Carbon Nanotubes with Nodal Structural Defects

*Young I. Jhon,[1] Woonjo Cho,[2] Seok Lee,[1] Young Min Jhon[1],\**

[1]Sensor System Research Center, Korea Institute of Science and Technology, Seoul 136-791, Republic of Korea
[2]Center for Opto-Electronic Conversion System, Korea Institute of Science and Technology, Seoul 136-791, Republic of Korea

**Abstract**

Recently experiments showed that nodal structural defects are readily formed in the synthesis of single-walled carbon nanotubes (SWNTs) and consequently, SWNTs are likely to deviate from well-defined seamless tubular structures. Here, using graphene-helix growth model, we describe structural details of feasible nodal defects in SWNTs and investigate how mechanical and electronic properties of SWNTs would change in the presence of them using computational methods. Surprisingly atomistic simulations of SWNTs with nodal defects show excellent agreement with previous structural, tensile, and ball-milling experiments whose results cannot be explained using conventional models. The tensile failure of SWNTs with nodal defects requires about four- or six-fold lower strength than pristine ones and these SWNTs are comparatively prone to damage under a lateral compressive biting. We reveal that electronic band-gap of SWNT(12,8) would be remarkably reduced in the presence of nodal defects. This study strongly indicates universality of nodal defects in SWNTs requesting new

\* *Corresponding author.* Tel: +82-2-958-5725. Email: ymjhon@kist.re.kr (Young Min Jhon)

theoretical framework in SWNT modelling for proper characteristics prediction.

Carbon nanotubes (CNTs) have drawn huge scientific and commercial interests from the discovery[1] due to their superb properties[2–4] and excellent compatibilities with other materials.[5-7] So far, enormous efforts have been placed on characterizing the structures and properties of these fascinating materials experimentally[8–10] and theoretically[11-13] in which most of theoretical works were carried out on the basis of well-defined seamless tubular structures of CNTs, possibly with several intrinsic topological point defects.[14,15]

During a long research history, however, remarkable discrepancies between experiments and theories of CNTs have been observed, which raise fundamental questions on their physical origins.[16–18] For instance, the tensile strength and yield strain of single-walled CNT (SWNT) was experimentally measured to be ~30 GPa and 5.3%, respectively, showing about four-fold deviation from the computation using seamless SWNT model.[16] In addition, the digitalized high-resolution transmission electron microscopy (HRTEM) images of Meyer et al.[17] showed atomically resolved nodal morphologies of SWNT but their physical origin and structural details have not been elucidated even considering any possible point defects in SWNTs.

Recently Lee et al.[19] observed the residual traces of the helices in SWNTs by HRTEM and atomic force microscopy (AFM) and suggested that the above puzzling experiments could be explained by the defects arising from graphene helix growth model. They supposed that CNT may grow, initially via a helical structure followed by "zipping" of the single strand, which distorts the graphene helix creating helical traces and the distorted graphene helix may have been previously interpreted as a chiral SWNT. They assumed that the atoms on edges of the unzipped zones of a helix may conceivably be "H-terminated" or structurally modified to minimize the conformation energy.

In this Letter, recognizing a possible, critical importance of nodal structural defects in SWNT characteristics, we attempt to identify the structures of feasible nodal defects in SWNT and examine how mechanical and electronic properties of SWNT would change in the presence of them using molecular dynamics simulations and density functional theory calculations.

We first modify some problematic points of Lee et al.'s graphene helix growth model as follows. In Lee et al.'s study, they assume that armchair-edged graphene nanoribbon would grow with the angle of ~60° to the substrate for the growth of SWNT. However, the magnitude of this growth angle is too large to be realized in practice. Furthermore, the growth of armchair-edged graphene nanoribbon is incompatible with experimentally observed chirality-dependent CNT population as indicated by Ding et al.[20] Therefore, we assume the growth of zigzag-edged graphene nanoribbon in this study.

We adopt graphene helix as structural units still but assume that graphene helix is generally perfectly zipped without distortion for many of its turns and it occasionally mismatches creating distinct nodal structural defects (Figure 1). We suppose that this lateral mismatch of graphene nanoribbon can possibly occur because the front growth of graphene nanoribbon seems faster than its lateral growth as inferred from experimentally observed spiral growth of SWNTs.[21] In our model, the axial distance between the nodes is determined by how many times the turn of graphene nanoribbon (with a rather small width) are completely matched. This model can effectively restrict nodal regions locally allowing for a large portion of seamless domains (Figure 1b) being compatible with many precedent studies showing normal

TEM images.

In designing nodal structures in SWNT, we consider two initial mismatched structures, one is as shown in Figure 1 and the other is done by allowing some overlap of graphene nanoribbon in order to assure a more chemically-broken nodal structure (Figure S1). Hereafter SWNTs that are incorporated with these nodal structures are denoted as SWNT-NOD-1 and SWNT-NOD-2, respectively.

Using these initial nodal structures, we obtain their equilibrated structures via geometric relaxation process followed by NPT and NVT ensemble simulations (see supporting information for details of the method). All calculations are performed using the LAMMPS (Large-scale Atomic/Molecular Massively Parallel Simulator) software package[22] and the simulation systems are constructed to meet a periodic boundary condition. For interactions between carbon atoms, we employed AIREBO (adaptive intermolecular reactive empirical bond order) potential[23] which has been widely used to study the mechanical properties of carbon materials.

The equilibrated structures of SWNT-NOD-1 and SWNT-NOD-2 are shown in Figure 2b and 2c, respectively, and we can see great similarities between our nodal structures and those of experimental TEM images (Figure 3).[17,18] It is noteworthy that "belly" structure is also observed in our model, similar to one of these TEM images.[17]

Next we have performed mechanical tensile processes of SWNT-NOD-1, SWNT-NOD-2, and their corresponding pristine SWNT (Figure2a, hereafter it is denoted as SWNT-PR), which has the same diameter and chirality of "(12,8)" to our nodes-incorporated SWNT systems, by using molecular dynamics simulation method (see supporting information). The tensile strength and yield strain of SWNT-PR are calculated to be 115.6 GPa and 17.1%, respectively, in good agreement with the reported values (94–123 GPa and 16.4–21.6%, respectively) of theoretical study using seamless SWNT model.[24] Meanwhile the tensile strength and yield strain of SWNT-NOD-1 are calculated to be 28.8 GPa and 4.3%, respectively, while they are amounted to be 21.0 GPa and 5.8% for SWNT-NOD-2, respectively. These tensile strengths of SWNT-NOD-1 and SWNT-NOD-2 are about four- or six-fold lower than that of SWNT-PR. However, they are in excellent agreement with the value (13–52 GPa) of Yu et al.'s experimental study[16] whose result has not been explained by conventional SWNT model so far.

The yield strains of CNT-NOD-1 (4.3%) and CNT-NOD-2 (5.8%) are also in excellent agreement with the value (5.3% or less) of Yu et al.'s study[16] which is much lower than the theoretical value (16.4–21.6%) obtained using seamless SWNT model. We also obtain reasonable Young's modulus compared to that in Yu et al.'s study.[16] The overall results obtained here are summarized in Table 1. Here it should be noted that mechanical fracture occurs at the weakest point of SWNTs and therefore, even in the case that nodal regions are incorporated into SWNTs with very low density, their effects are fatal to tensile fracture. In other word, seemly perfect SWNTs may possess nodal defects and thus can exhibit degraded mechanical performance, which might be more usual in SWNTs than we expect.

In our tensile simulations, the fracture point is determined to be the point that exhibits the maximum tensile stress followed by its subsequent prestigious drop (Figures S2−S4). In fact we observe fluctuating residual stresses even after this tensile failure due to the connection via graphene nanoribbons that come from unraveling of nodal defect structure (Figure S5).

This residual response could be observable in the samples with nm-scale length like our simulation systems. However it will be hardly captured in the systems using μm-scale samples, which is the same level as the most experimental systems including Yu et al.'s one,[16] because such inhomogeneous deformation emerges very shortly in μm-scale samples due to the extremely large ratio ($10^3$−$10^4$) of a sample length to a nodal region length. We suppose that the trace of graphene nanoribbons can possibly be shown at the fracture section even for μm-scale CNTs and suggest experimentalists to carefully probe tensile fracture section structures of commercial arc-SWNTs which may distinctly differ from that of seamless tubular one.

Inspired by graphene helix growth model, Lee et al. have performed an exquisite ball-milling process on (multi-walled) CNTs to dissolve these CNTs into multiples of graphene nanoribbons.[25] In that study, even after a short ball-milling process, they can obtain severely destroyed structures of CNTs (Figure S6), which are far from the structures we can expect from seamless tubular CNTs. To investigate the role of nodal defects in the ball-milling process, we have performed atomistic simulations for lateral compressive biting of our SWNT systems (Figure 4). We find that the structures of SWNT-NOD-1 and SWNT-NOD-2 critically damages (especially for SWNT-NOD-2) after the process while SWNT-PR can sustain the structure quite fairly (Figure 5) supporting the conjecture on universality of nodal defects in SWNTs. To manifest this fact, we examine their hysteresis loops of internal energy which show that the process is rather reversible for SWNT-PR while remarkable increases of internal energy are observed for SWNT-NOD-1 and SWNT-NOD-2 after one cycle of the biting process (Figure 6a). Despite such critical damages, we find that the maximum forces of the indenters required for SWNT-NOD-1 and SWNT-NOD-2 are smaller than that for SWNT-PR although the magnitudes of their indenter forces are generally similar to one another during the process (Figure 6b). The structural variations of our SWNT systems observed during the overall period of the process are shown in Figures S7−9 and movies S1−3.

Finally we have investigated how electronic structure of SWNT(12,8) (i.e., SWNT-PR) would change in the presence of nodal defects using density functional theory calculations (see supporting information for details of the method) as embedded in ATK software package.[26] The analyses on electron band structures (Figure 7) and density of states (Figure S10) indicate that electron energy-gap of SWNT(12,8) would drastically narrows from 0.482 eV to 0.156 and 0.0837 eV, respectively, if nodal defects pertaining to SWNT-NOD-1 and SWNT-NOD-2 are created with the density of ~0.027/Å . The energy-gap narrowing could be weakened as the nodal defect density decreases. We suppose that this energy-gap narrowing is attributed to the inclusion of metallic zigzag-edged graphene nanoribbons present in the nodal region.

In summary, on the basis of graphene helix growth model, we have proposed structural details of feasible nodal defects in SWNTs, which are consistently compatible with experimentally observed TEM images. The atomistic simulation of SWNTs with nodal defects excellently reproduces previous tensile experiment whose results greatly differ from those of seamless SWNT and its physical origin has been elusive for a long time. We demonstrate that electronic structures of SWNTs could remarkably change (toward metallic characteristics) in the presence of nodal defects. This study strongly indicates universality of nodal defects in SWNTs and requests new theoretical framework in SWNT modelling for proper characteristics prediction.

**Acknowledgements**

This work was supported by "Industrial Strategic Technology Development Program (10039226, Development of actinic EUV mask inspection tool and multiple electron beam wafer inspection technology) funded by the Ministry of Trade, Industry & Energy, Republic of Korea", "the KIST Institutional Program (Project No. 2E24572)", and "Global Top Project (GT-11-F-02-002-1) funded by the Korea Ministry of Environment".

**Supporting Information**

Computational details, structural figures and movies for further understanding, and tensile stress-strain curves.

**Table 1.** Tensile strength, yield strain, and elastic modulus of SWNTs obtained from experiment, conventional seamless model, and our node-based model.

|  | Stress (GPa) | Strain | Elastic Modulus (GPa) |
| --- | --- | --- | --- |
| SWNT-PR | 115.6 | 0.1712 | 960.8 |
| SWNT-NOD-1 | 28.8 | 0.0426 | 651.2 |
| SWNT-NOD-2 | 21.0 | 0.058 | 367.0 |
| SWCNT(9,0) (theory)[24] | 94.0 | 0.164 | 939.1 |
| SWCNT(5,5) (theory)[24] | 123.0 | 0.216 | 894.7 |
| SWCNT (experiment)[16] | 13–52 (mean:30) | < 0.053 | 320–1470 (mean:1006) |

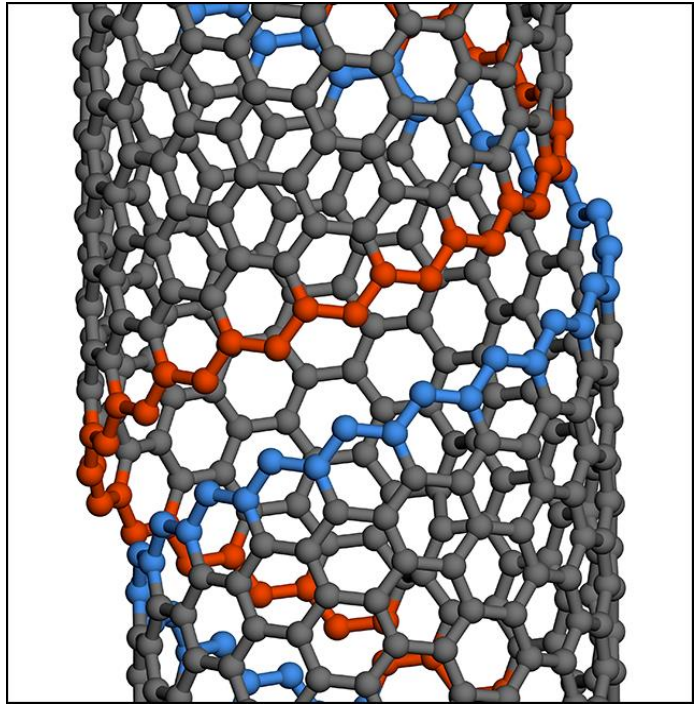

(a)

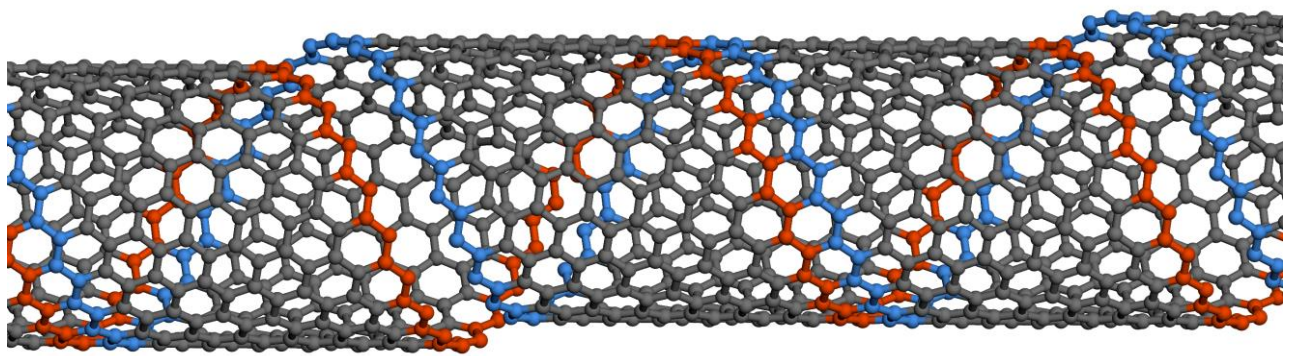

(b)

**Figure 1.** The mismatched graphene helix in SWNT **(a)** and the overall view of its associated SWNT **(b)**. Blue and red indicate two opposite edges of graphene nanoribbon.

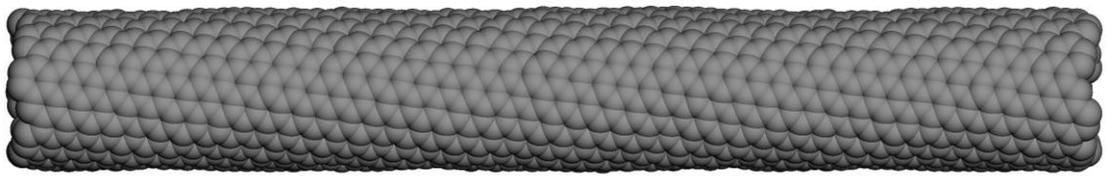

(a)

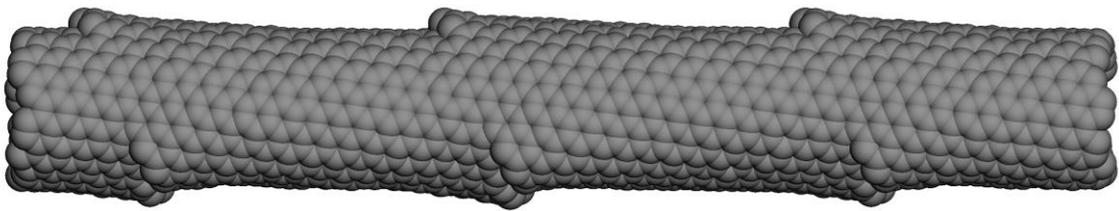

(b)

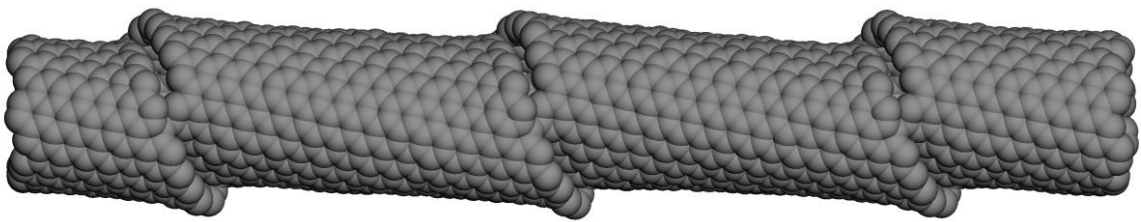

(c)

**Figure 2.** The equilibrated structures of SWNT-PR **(a)**, SWNT-NOD-1 **(b)**, SWNT-NOD-2 **(c)**.

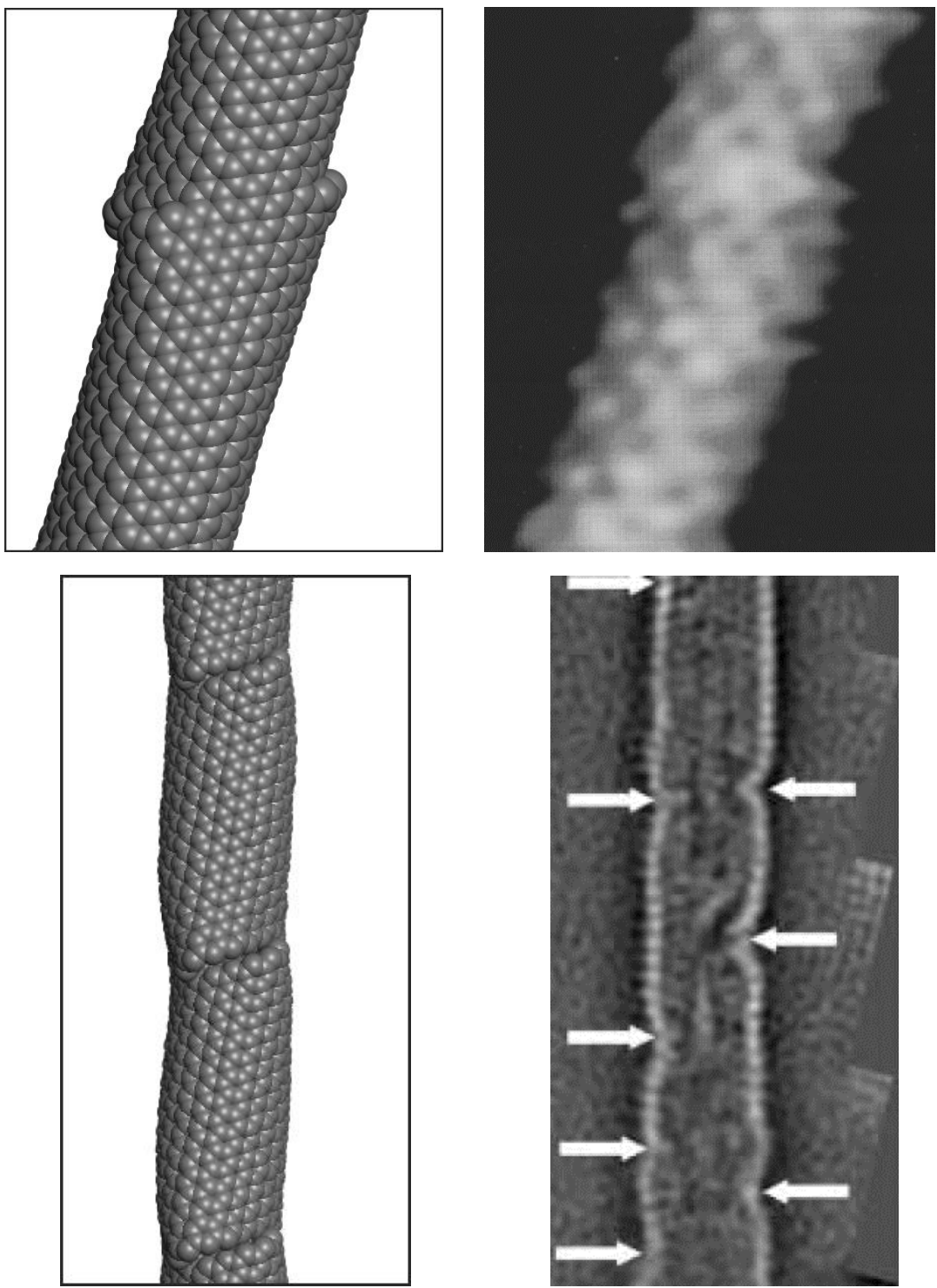

**Figure 3.** The structures of SWNT-NOD-1 (left column) and the TEM images of SWNTs in experiments (right column).[17,18]

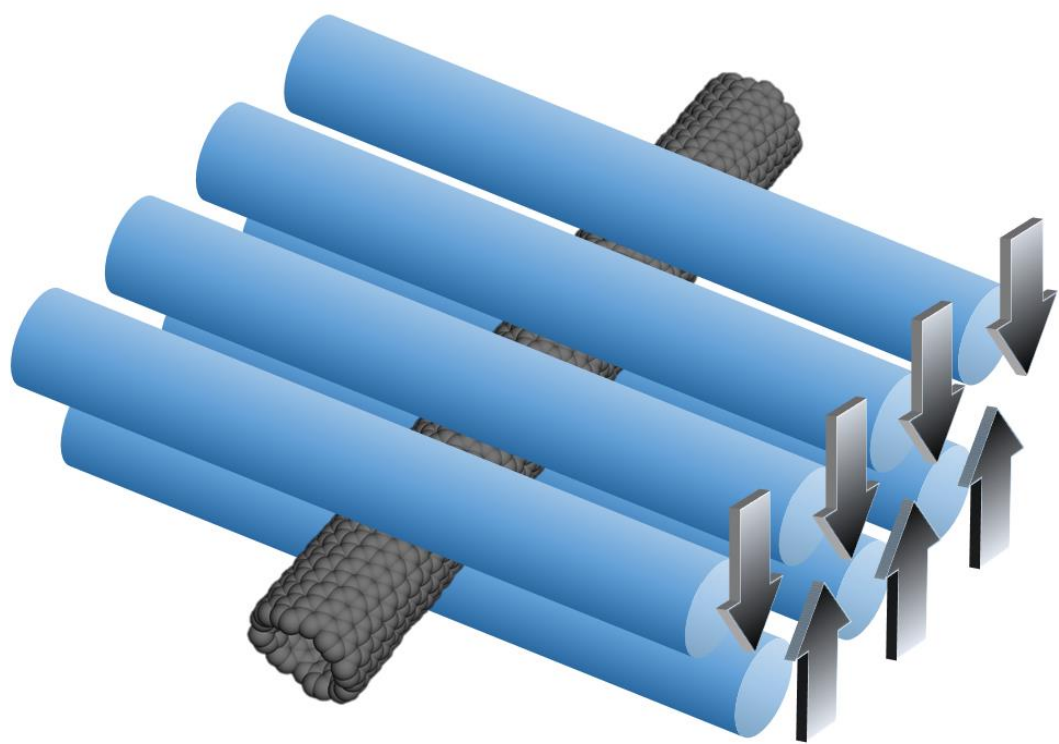

**Figure 4.** Schematic of lateral compressive biting of SWNTs designed to simulate a ball-milling process.

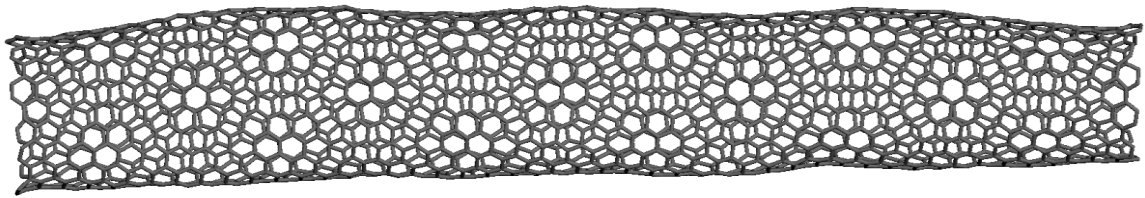

(a)

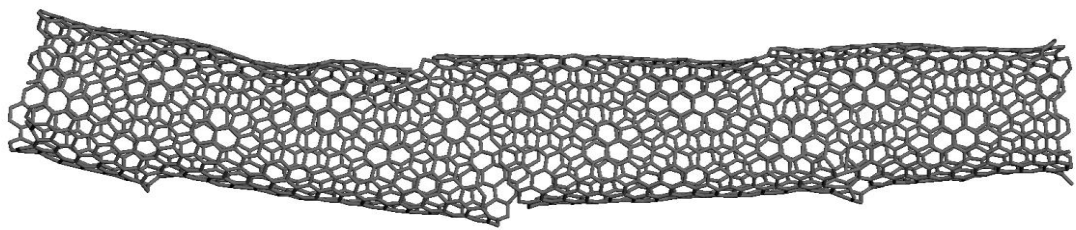

(b)

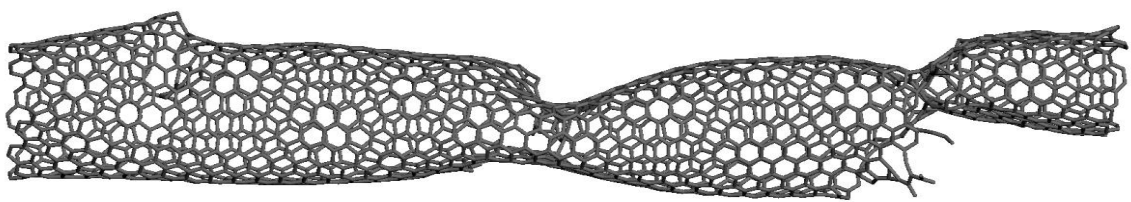

(c)

**Figure 5.** The structures of SWNT-PR **(a)**, SWNT-NOD-1 **(b)**, SWNT-NOD-2 **(c)** obtained after one cycle of lateral biting process.

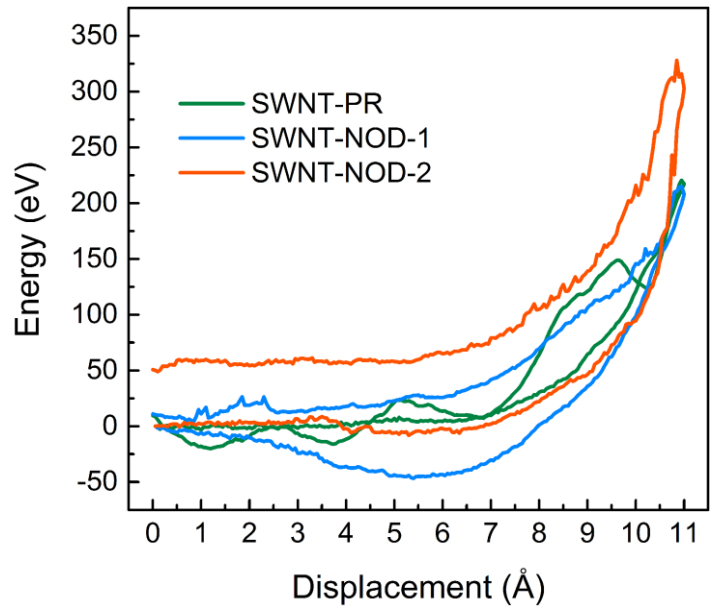

(a)

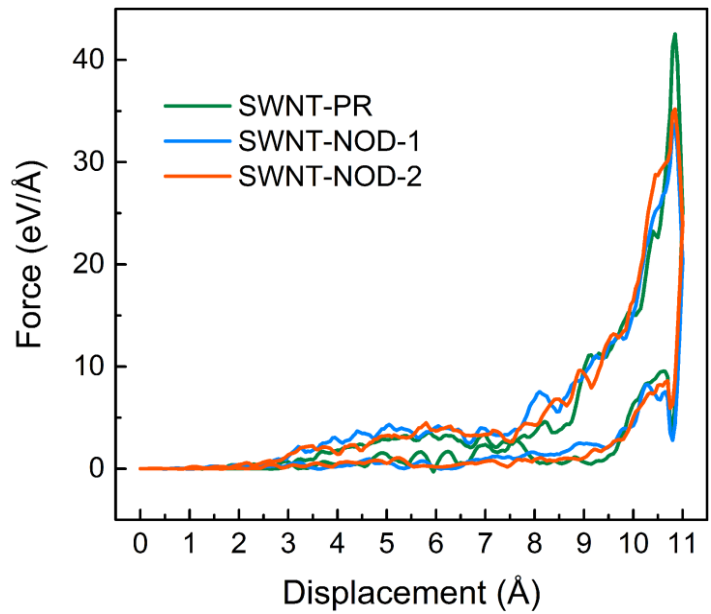

(b)

**Figure 6.** The energy hysteresis loops (counterclockwise) **(a)** and the variations of the indenter force (clockwise) **(b)** observed during the lateral biting processes of our SWNT systems.

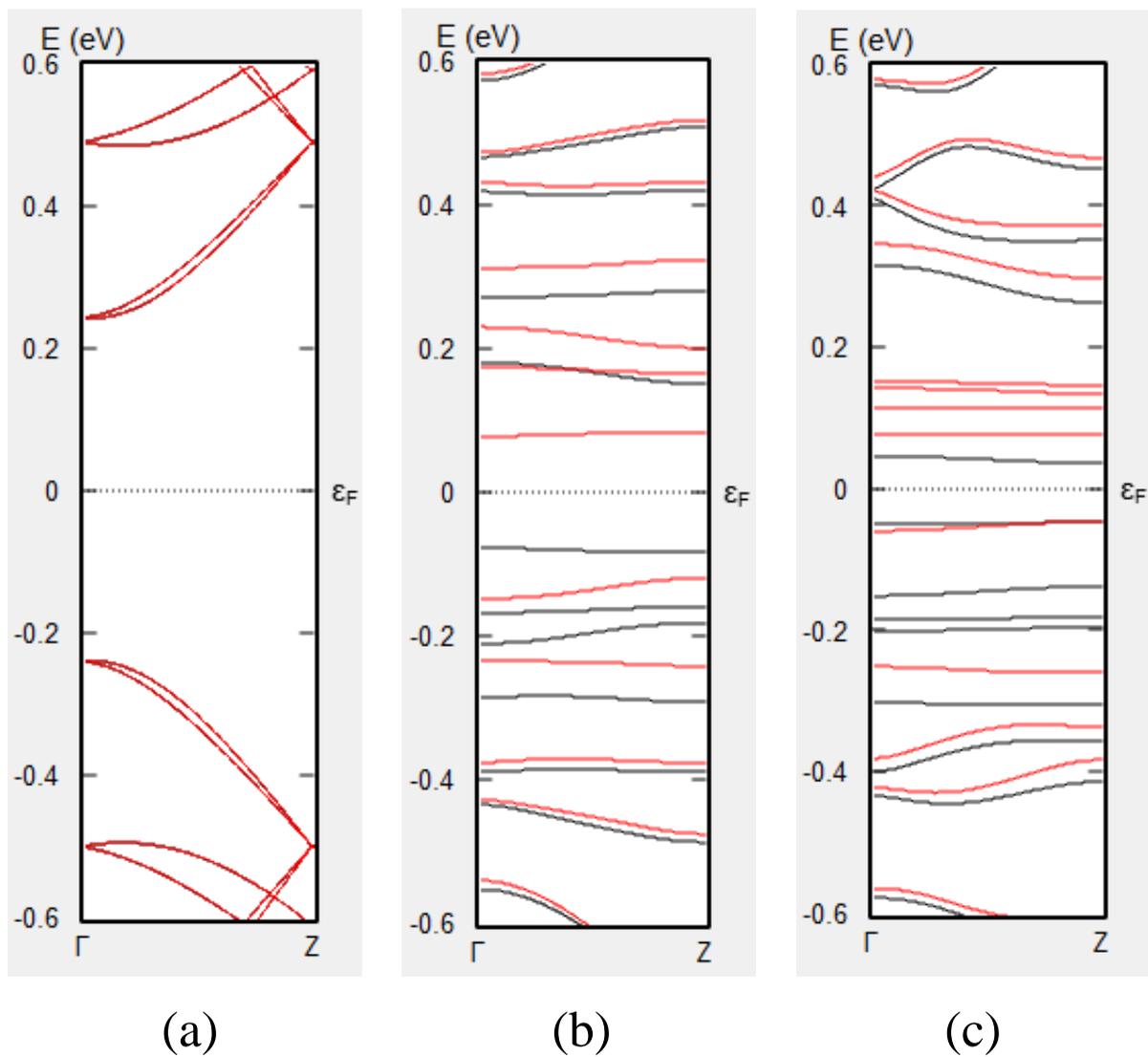

**Figure 6.** Electron band structures of SWNT-PR **(a)**, SWNT-NOD-1 **(b)**, and SWNT-NOD-2 **(c)**. Black and red indicate different spin contributions.